\documentclass[sn-mathphys-num]{sn-jnl}

\usepackage{graphicx,amsmath,amssymb,enumerate}
\usepackage{bm,hyphenat,xspace} \usepackage{graphicx,epsfig}

\newcommand {\mbf}[1]{{\mathbf{#1}}} 

            \newcommand{\Ca}{{}^{12}\mathrm{C}}
           \newcommand{\Ox}{{}^{12}\mathrm{O}}
           \newcommand{\Mg}{{}^{12}\mathrm{Mg}}

\newcommand{\cm}{\mathrm{c\!\:\!.m\!\:\!.}}

\begin{document}

\title { Three-body calculation of deuteron-nucleus scattering using microscopic global optical potential}

\author{A. Deltuva}
\author{D. Jur\v{c}iukonis}
\author{D.~Likandrovas}
\author{J.~Torres~Fernandez}

\affil{Institute of Theoretical Physics and Astronomy, Vilnius University, LT-10222 Vilnius, Lithuania}


\abstract{
  We test microscopic global optical potential in three-body calculations
  of deuteron-nucleus scattering. We solve Faddeev-type equations for three-body
  transition operators. We calculate differential cross section
and analyzing power  
for the deuteron elastic scattering and breakup in collisions with
  ${}^{12}$C, ${}^{16}$O and ${}^{24}$Mg nuclei, and find a reasonable agreement
  with available experimental data. Comparison with respective predictions
  using phenomenological optical potentials reveals systematic deviations in particular
  kinematic regimes.}

 \maketitle

\section{ Introduction} 

Optical potentials are essential quantities in nuclear reaction theory
enabling the reduction of a two-cluster  many-nucleon scattering problem to an effective two-body problem.
Due to inability to solve the many-body problem in the continuum even approximately, the construction 
of the optical potentials relied on the phenomenology for a long time.
Over years  a number of quite successful nucleon-nucleus interaction models have been developed
by adjusting their parameters to the experimental data,
examples being the optical potentials by Becchetti and Greenlees \cite{becchetti},
Watson et al.~\cite{watson}, Menet et al.~\cite{menet},
Perey and Buck \cite{pereybuck}, Giannini and Ricco \cite{giannini}, Chappel Hill \cite{CH89},
Koning and Delaroche \cite{koning}, Weppner et al. \cite{weppner:op}, and many others.
Last decades witnessed an enormous progress in many-nucleon structure calculations using various methods,
that propagated into the microscopic calculations of the optical potentials with the
application of Watson multiple scattering, no-core shell model with continuum,
self-consistent Green's function,  nuclear matter and other approaches;
see Ref.~\cite{microOP:23} for a review. As pointed out in Ref.~\cite{microOP:23},
most of those optical potentials are nonlocal and have no analytical representation which
precludes them 
from 
being widely used in few-body nuclear reaction calculations.
In this sense the work by Furumoto et al. \cite{furumoto:mgop}
is rather an exception since it provides a user-friendly code for
 the microscopic global optical potential (MGOP).
 This potential relies on  a single-folding model with the complex $G$-matrix interaction.
 The single-folding model employs nuclear densities that are calculated using two microscopic
 mean-field models: the relativistic-mean-field (RMF) and Skyrme-Hartree-Fock + BCS (HF+BCS) models,
 the resulting microscopic potentials are abbreviated by MGOP1 and MGOP2, respectively.
It is developed for even-even nuclei encompassing mass numbers from 10 to 276,
 both stable such as $\Ca$ and neutron- or proton rich such as ${}^{22}\mathrm{C}$ or
 ${}^{12}\mathrm{O}$, and is valid from 50 to 400 MeV nucleon beam energy.
 The potential consists of real central, imaginary central, real spin-orbit, and imaginary spin-orbit parts,
that  calculated microscopically but for further applications expanded into a sum of 12 Gaussians.
As shown in Ref.~\cite{furumoto:mgop}, the resulting shape may deviate from the typical
Woods-Saxon form most often employed in phenomenological optical potentials
 \cite{becchetti,watson,menet,CH89,koning}. Nevertheless,  the 
experimental nucleon-nucleus scattering data is reproduced with a comparable quality.
It is therefore interesting to compare the 
prediction 
of MGOP and traditional phenomenological
optical potentials in more complicated few-body nuclear reactions. Among them elastic deuteron-nucleus scattering
and deuteron breakup is the simplest nontrivial case. Calculation of these collision processes
requires the nucleon-nucleus interaction in continuum only. In contrast, for the deuteron
stripping and pickup also binding potentials would be needed that are not provided by the MGOP approach.

We therefore will study deuteron elastic scattering and breakup in the  collision with several
light nuclei as a three-body problem. Most often these processes are described using  the 
continuum-discretized coupled channels (CDCC) method  \cite{austern:87}. The method
has been benchmarked in Ref.~\cite{deltuva:07d}
against the rigorous Faddeev scattering theory \cite{faddeev:60a,alt:67a} and
confirmed to be reliable for deuteron elastic scattering and breakup at 56 MeV deuteron beam energy.
As the benchmark at higher energies was not performed, and the MGOPs are applicable above 50 MeV/nucleon energy,
our study will be based on rigorous Faddeev-type equations. We shortly recall our theoretical formalism
in Sec. 2, while in Sec. 3 we present results for the deuteron scattering off $\Ca$, $\Ox$ and $\Mg$ nuclei
in the 100 - 200 MeV beam energy region. Summary is given in Sec. 4.

\section{Alt, Grassberger and Sandhas equations} 

We describe the deuteron-nucleus scattering using the
Alt, Grassberger, and Sandhas (AGS) equations  \cite{alt:67a}
 for three-particle transition operators
\begin{subequations}  \label{eq:AGScpl}   
\begin{align}  
  U_{AA} = {}& t_p G_0 U_{pA} + t_n G_0 U_{nA} , \\
U_{pA} = {}& G_0^{-1} +  t_A G_0 U_{AA} + t_n G_0 U_{nA}, \\
U_{nA} = {}& G_0^{-1} +  t_A G_0 U_{AA} + t_p G_0 U_{pA},
\end{align}
\end{subequations}
where Latin subscripts label the spectator particle (and simultaneously also the pair
in the  odd-man-out notation), being either nucleus ($A$), proton ($p$) or neutron ($n$).
The initial state subscript $A$ indicates the initial configuration of bound pair of nucleons and 
free spectator nucleus $A$. All three-body transition operators in Eqs.~(\ref{eq:AGScpl})
act on the corresponding channel state which 
is a product of a
 deuteron bound-state wave function $|\phi_d \rangle$ and a free wave $|\mbf{q}_A \rangle$
for the  nucleus-deuteron relative motion. 
Furthermore, 
\begin{equation}  \label{eq:g0}   
 G_0 =  (E+i0  - H_0)^{-1}
\end{equation}
is the free resolvent at
the available energy $E$ in the center-of-mass (c.m.) frame, while $H_0$ is the kinetic energy operator
for the relative motion of three particles.
The potentials $v_a$ for all pairs $a$ 
do not enter the AGS equations directly but via the two-particle transition operators
\begin{equation}  \label{eq:ta}   
 t_a =   v_a +  v_a G_0 t_a.
\end{equation}
The breakup operator 
\begin{equation}  \label{eq:AGS0}   
U_{0A} =  G_0^{-1} + t_A G_0 U_{AA} + t_p G_0 U_{pA} + t_n G_0 U_{nA} 
\end{equation}
is obtained by integration once the AGS equations (\ref{eq:AGScpl}) for other transition operators are solved.
The final breakup state is a product of two free waves for the relative motion of three particles.
The matrix elements of  $U_{AA}$ and  $U_{0A}$ taken between the respective initial and final
channel states yield the transition amplitudes for elastic deuteron scattering and breakup, respectively.
The AGS equations are solved in the momentum-space;
further details of calculations, also with respect to the inclusion of the proton-nucleus Coulomb
force via the method of screening and renormalization \cite{taylor:74a,semon:75a,alt:80a,deltuva:05a},
can be found in Ref.~\cite{deltuva:07d} and references therein.

 The MGOP potentials by Furumoto et al. \cite{furumoto:mgop} are given as local potentials in the
 coordinate-space, thus, we perform the transformation to the momentum-space numerically in a standard way.
 In addition to MGOP1 and MGOP2 we perform calculations using phenomenological optical potentials
 by Koning and Delaroche (KD) \cite{koning} and Weppner et al. \cite{weppner:op}. The latter is one of rather
 few global potentials fitted to light nuclei such as $\Ca$ and $\Ox$ in a rather broad energy range
 30 to 160 MeV. The KD potential is fitted to $\Mg$ and heavier nuclei in the 1 keV to 200 MeV regime,
 but is being used in many studies also for lighter nuclei, and found to provide a reasonable description.
 As customary in the deuteron scattering, we take optical potentials with parameters evaluated
 at half the deuteron beam energy. For the neutron-proton interaction we use the high-precision
 CD Bonn potential \cite{machleidt:01a}; we admit that the sensitivity of the results to
 the version of a realistic  neutron-proton force is very low.

\section{Deuteron elastic scattering}

\begin{figure}[!]
\begin{center}
\includegraphics[scale=0.4]{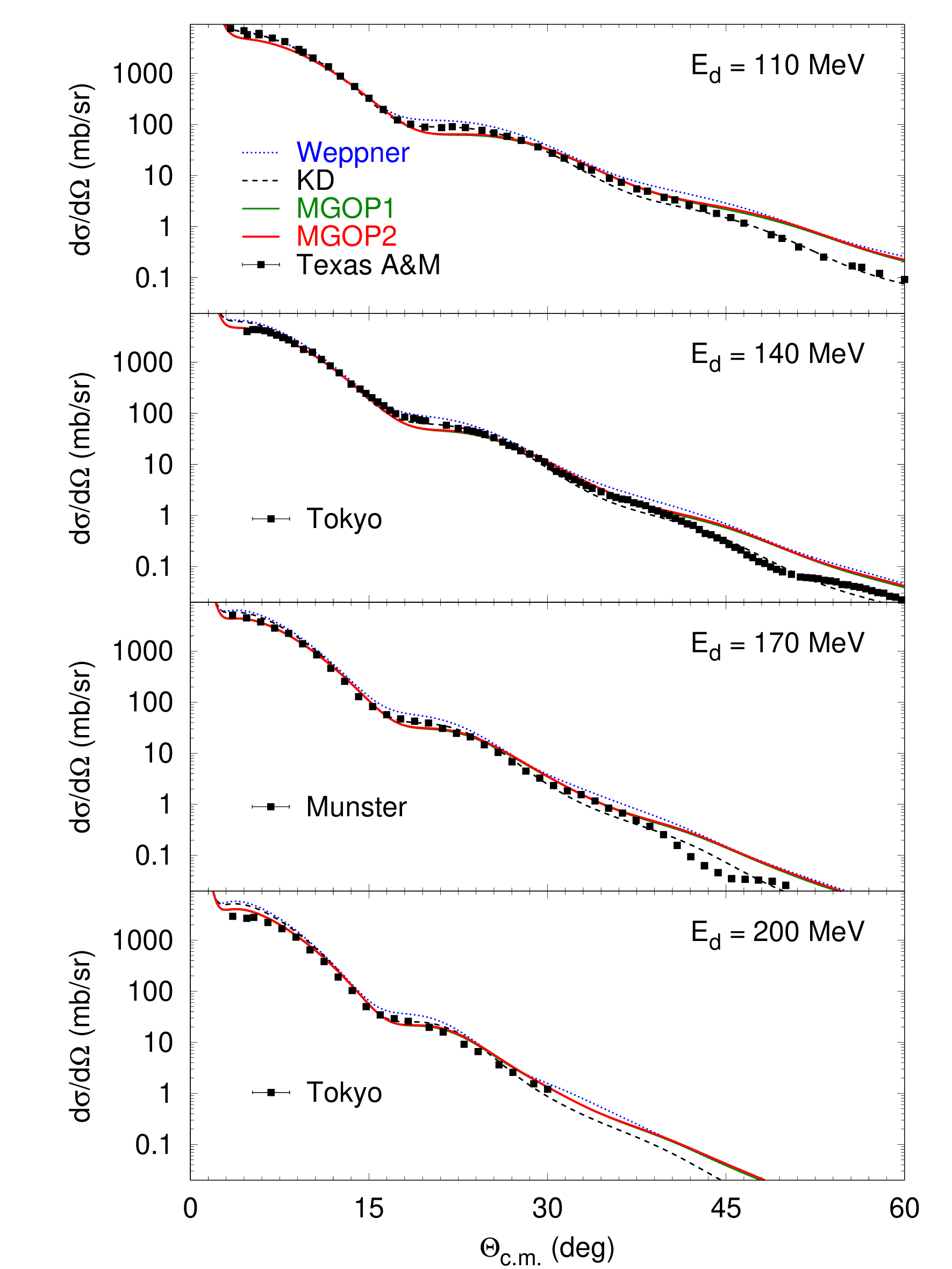}    
\end{center}
\caption{\label{fig:cs12C} Differential cross section for $d$ + $^{12}$C elastic scattering at deuteron beam energies
 $E_{d}$ = 110, 140, 170 MeV and 200 MeV as function of the c.m. scattering angle $\Theta_{\mathrm{c.m.}}$. 
Results obtained with four optical potentials are compared with experimental data from 
Refs.~\cite{PhysRevC.48.2085}, \cite{PhysRevC.58.2180}, \cite{PhysRevC.63.037601} and \cite{PhysRevC.70.034318}.}    
\end{figure}

In Fig.~\ref{fig:cs12C} we show the angular dependence of the differential cross section for
the elastic deuteron-$\Ca$ scattering at deuteron beam energies $E_d =110$ 140, 170 and 200 MeV. 
All considered optical potentials provide quite reasonable description of the experimental data 
\cite{PhysRevC.48.2085}, \cite{PhysRevC.58.2180}, \cite{PhysRevC.63.037601} and \cite{PhysRevC.70.034318}
but no one of them accounts for the data perfectly in the whole energy and angle regime.
Remarkably, although the KD potential is fitted to $\Mg$ and heavier nuclei, among all considered potentials 
the predictions of KD best describe  the experimental data for elastic deuteron-$\Ca$ scattering at larger angles 
$\Theta_{\mathrm{c.m.}}$ above 40 deg.
The predictions of MGOP1 and MGOP2 are almost indistinguishable, and are slightly lower 
than the ones of Weppner and KD potentials  at forward angles below 10 deg.

\begin{figure}[!]
\begin{center}
\includegraphics[scale=0.4]{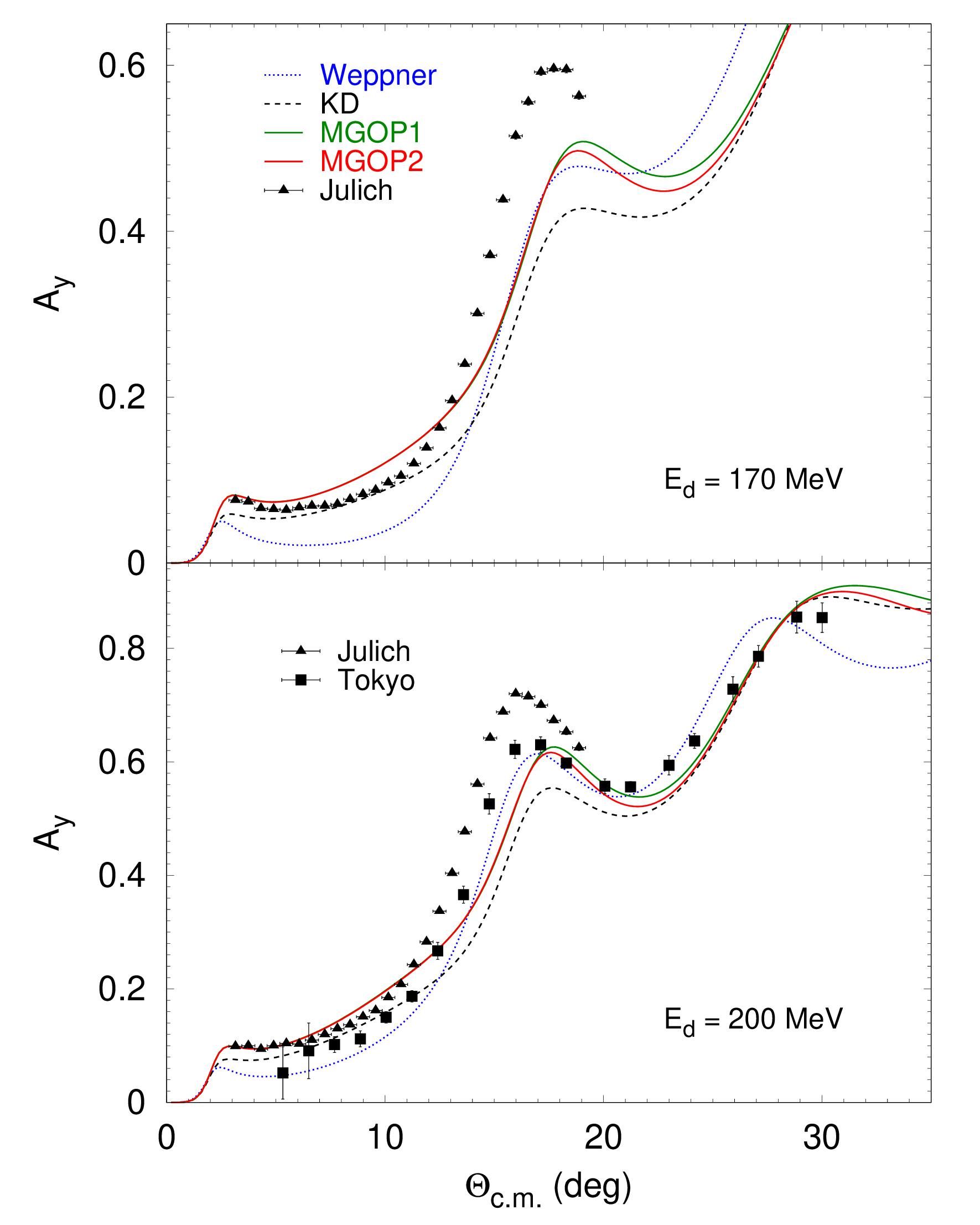}    
\end{center}
\caption{\label{fig:anapow12C} Deuteron vector analyzing power for $d$ + $^{12}$C  elastic  scattering at $E_{d}$ = 170 and 200 MeV. 
Results obtained with four optical potentials are compared with experimental data from 
Refs.~\cite{Eur.Phys.J.56} and \cite{PhysRevC.70.034318}.}    
\end{figure}

In Fig.~\ref{fig:anapow12C} we study the deuteron vector analyzing power for $d$ + $^{12}$C  elastic  scattering 
at $E_{d}$ = 170 and 200 MeV. 
All four optical potentials roughly reproduce the shape of the angular dependence but underpredict the 
data from Ref.~\cite{Eur.Phys.J.56} around the peak near $\Theta_{\mathrm{c.m.}}=15$ deg.
On the other hand, at $E_{d}$ = 200 MeV the data from Ref.~\cite{PhysRevC.70.034318} are accounted for 
considerably better, thus, some concerns regarding the compatibility and reliability of the data should be raised.
Again, the predictions of MGOP1 and MGOP2 are much closer to each other than any other potentials.

\begin{figure}[h!]
\begin{center}
\includegraphics[scale=0.75]{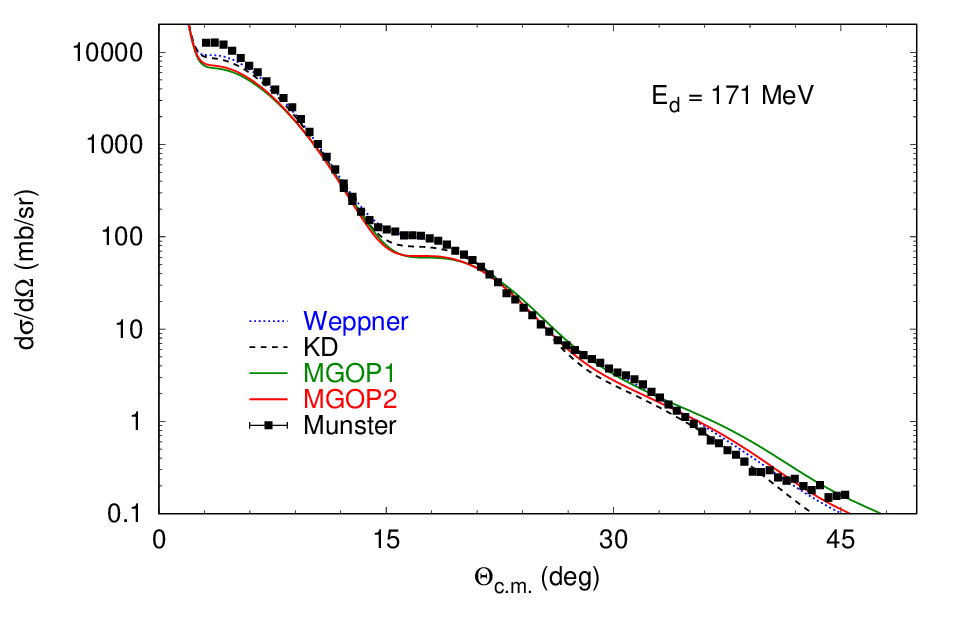}    
\end{center}
\caption{\label{fig:csO171} Differential cross section for $d$ +$^{16}$O elastic scattering at  $E_{d}$ = 171 MeV.
 Results obtained with four optical potentials are compared with experimental data from Ref.~\cite{PhysRevC.70.067601}.}
\end{figure}

\begin{figure}[!]
\begin{center}
\includegraphics[scale=0.75]{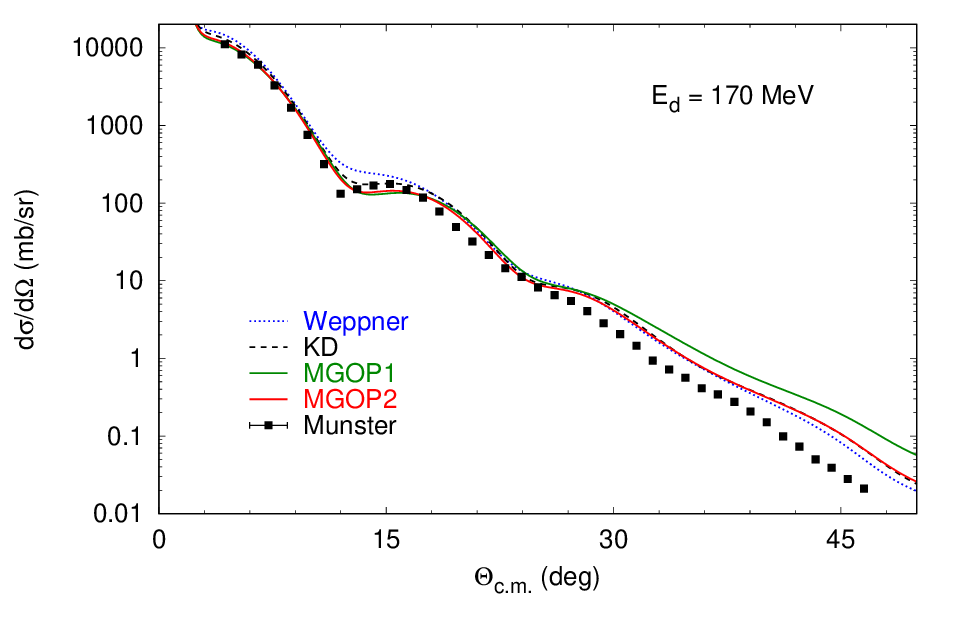}    
\end{center}
\caption{\label{fig:csMg170} Differential cross section for   $d$ +$^{24}$Mg elastic scattering at $E_{d}$ = 170 MeV.
Results obtained with four optical potentials are  compared with experimental data from Ref.~\cite{PhysRevC.63.037601}.}    
\end{figure}

The results of the differential cross section for $d$ +$^{16}$O elastic scattering at  $E_{d}$ = 171 MeV are shown in Fig.~\ref{fig:csO171}.
This time it is the Weppner potential that reproduces the data \cite{PhysRevC.70.067601} best, 
while both MGOP models underpredict the cross section at forward angles and at the minimum near 15 deg.
On the contrary, the differential cross section for $d$ +$^{24}$Mg elastic scattering at $E_{d}$ = 170 MeV \cite{PhysRevC.63.037601}
shown in Fig.~\ref{fig:csMg170}  is better reproduced by MGOP models, while the one by Weppner leads to an overprediction.
At larger beyond 30 deg also moderate differences between
 MGOP1 and MGOP2 predictions can be seen.

\section{Semi-inclusive deuteron breakup on $\Ca$}

\begin{figure}[!]
\begin{center}
\includegraphics[scale=0.5]{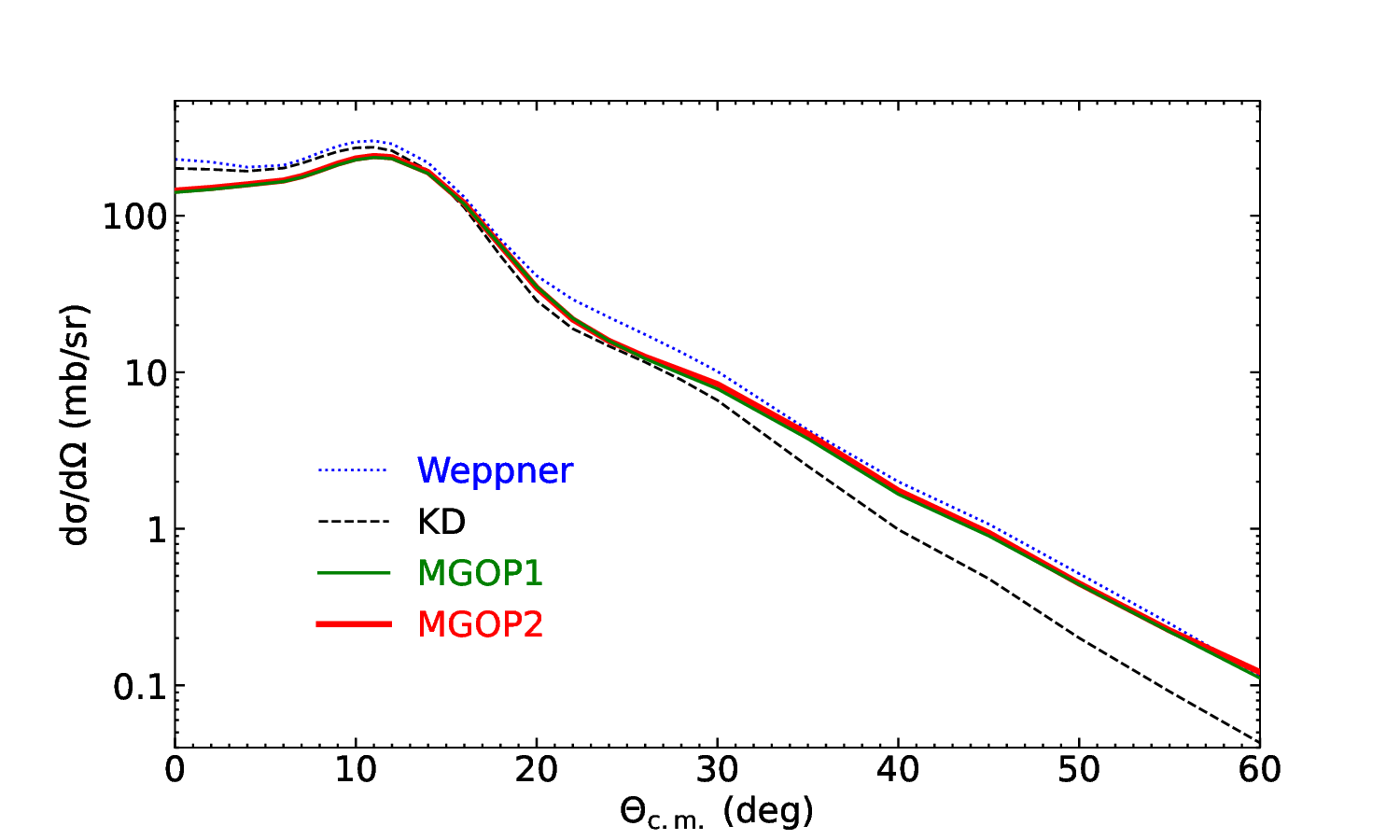}
\end{center}
\caption{\label{fig:ba} (Color online) Semi-inclusive
differential cross section for the deuteron breakup on the ${}^{12}$C nucleus
at  $E_d = 140$  MeV as a function of the  ${}^{12}$C scattering angle $\Theta_{\cm}$
in the  c.m. frame. Predictions obtained with four optical potentials are compared.
}
\end{figure}

\begin{figure}[!]
\begin{center}
\includegraphics[scale=0.5]{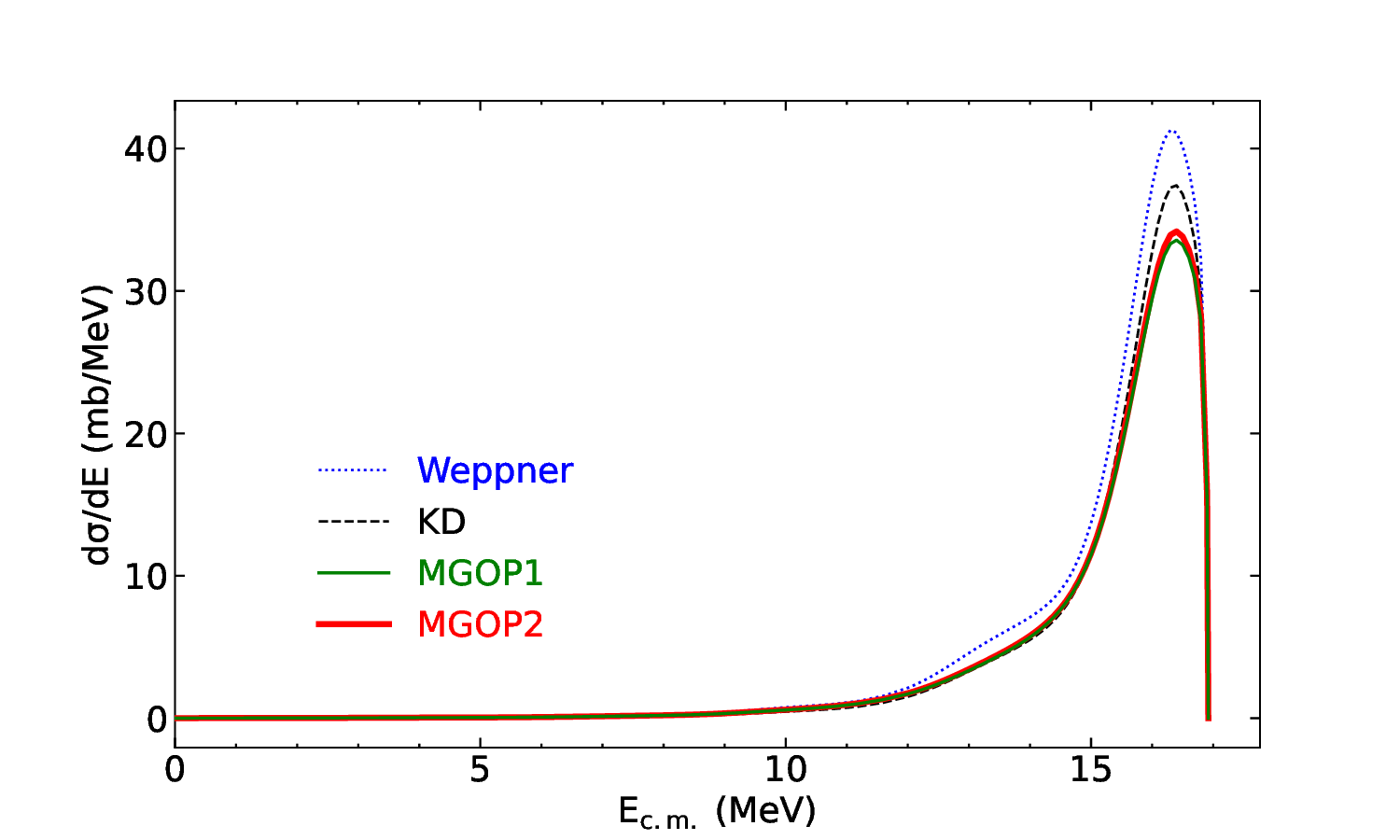}
\end{center}
\caption{\label{fig:be} (Color online) Semi-inclusive
differential cross section for the deuteron breakup on the ${}^{12}$C nucleus
at  $E_d = 140$  MeV as a function of the  ${}^{12}$C energy $E_{\cm}$
in the  c.m. frame. Predictions obtained with four optical potentials are compared.
}
\end{figure}

In Figs.~\ref{fig:ba} and \ref{fig:be}
we show an example for angular and energy distributions of the semi-inclusive
differential cross section for the deuteron breakup in collision with  $\Ca$ nucleus.
The beam energy is 140 MeV, and we assume  $\Ca$ to be the detected particle.
Predictions using MGOP1 and MGOP2 are very similar and systematically lower than
those of KD and Weppner potentials under the dominating kinematic conditions, i.e.,
small scattering angles and nearly maximal allowed energy. The latter condition corresponds
to the vanishing relative neutron-proton energy, with an enhancement due to the virtual ${}^1S_0$ state.
At scattering angles above $30^\circ$ the predictions using KD potential become lower
than for other potentials, resembling also the situation in the elastic scattering. 
This could be a consequence of not constraining the KD
potential by the nucleon-$\Ca$ data. On the other hand, for the energy distribution it
is the Weppner potential that differs most from the rest.

\section{Summary and conclusions}

We performed three-body calculations for deuteron elastic scattering and breakup in the
collision with a nucleus. We considered  $\Ca$, $\Ox$ and $\Mg$ nuclei and
used several phenomenological as well as microscopic global optical potentials.
We obtained a reasonable reproduction of the experimental data 
for the elastic differential cross section and deuteron vector analyzing power,
although no one of the potentials is able to account for the data perfectly in the whole
energy and angle regime.
Under particular kinematic conditions,
most notably at forward angles in both elastic deuteron scattering and breakup,
 the predictions using MGOP  deviate
from those of phenomenological potentials.
In general, our work provides further support for the applicability
of MGOP in the description of few-body reactions.

\vspace{1mm}
  {\bf Acknowledgments.} 
This work was supported by Lietuvos Mokslo Taryba
(Research Council of Lithuania) under Contract No.~S-MIP-22-72.


\end{document}